\documentclass[a4paper,
              ]{jacow}

\makeatletter%                           % test for XeTeX where the sequence is by default eps-> pdf, jpg, png, pdf, ...
\ifboolexpr{bool{xetex}}                 % and the JACoW template provides JACpic2v3.eps and JACpic2v3.jpg which might generates errors
 {\renewcommand{\Gin@extensions}{.pdf,%
                    .png,.jpg,.bmp,.pict,.tif,.psd,.mac,.sga,.tga,.gif,%
                    .eps,.ps,%
                    }}{}
\makeatother

\ifboolexpr{bool{xetex} or bool{luatex}} % test for XeTeX/LuaTeX
 {}                                      % input encoding is utf8 by default
 {\usepackage[utf8]{inputenc}}           % switch to utf8

\usepackage[USenglish]{babel}

\ifboolexpr{bool{jacowbiblatex}}%        % if BibLaTeX is used
 {%
  \addbibresource{jacow-test.bib}
  \addbibresource{biblatex-examples.bib}
 }{}

\begin{document}

\title{Study of Phase Reconstruction Techniques applied to Smith-Purcell Radiation Measurements\thanks{Work supported by the French ANR (contract ANR-12-JS05-0003-01), the PICS (CNRS) "Development of the instrumentation for accelerator experiments, beam monitoring and other applications and  Research Grant \#F58/380-2013 (project F58/04) from the State Fund for Fundamental Researches of Ukraine in the frame of the State key laboratory of high energy physics." }}

\author{
N. Delerue~\thanks{delerue@lal.in2p3.fr}, J. Barros,  M. Vieille-Grosjean, LAL, Orsay, France\\
Oleg Bezshyyko, Vitalii Khodnevych, Taras Shevchenko National University of Kyiv , Ukraine}

\maketitle

\begin{abstract}
Measurements of coherent radiation at accelerators typically give the absolute value of the beam profile Fourier transform but not its phase. Phase reconstruction techniques such as Hilbert transform or Kramers Kronig reconstruction are used to recover such phase. We report a study of the performances of these methods and how to optimize the reconstructed profiles. 
\end{abstract}

\section{Longitudinal bunch profile measurement at particle accelerators}

On a particle accelerator the longitudinal profiles of a particle bunch can not easily be measured. Several indirect measurement techniques have been established relying on the radiation emitted by the bunch either when it crosses a different material~\cite{OTR_LURE} or when it passes near a different material~\cite{ODR_Cianchi,Doucas_ESB}. In this case what is measured is the emitted spectrum. This emitted  spectrum encode the longitudinal profile through the relation:

\begin{equation}
I(\lambda) = I_1(\lambda) ( N + | F(\lambda) |^2 N^2 )
\end{equation}

where $I(\lambda)$ is the emitted intensity as a function of the wavelength.  $I_1(\lambda)$ is the intensity of the signal emitted by a single particle and $F(\lambda)$ is a form factor that encodes the longitudinal and transverse shape of the particle bunch. Recovering the longitudinal profile requires to invert this equation however this is not straightforward as the information about the phase of the form factor can not be measured and therefore is not available.

A phase reconstruction algorithm must therefore be used to recover this phase. Several methods exist (see for example~\cite{KK}). We have implemented two of these methods and we assess their performances below.

\section{Reconstruction methods}

When it is only possible to measure the amplitude of the complex signal, it is necessary to recover the phase of the available data. 
For an analytic function this is easier because the real and imaginary part are not completely independent.
The Kramers-Kronig relations~\cite{KK} helps restore the imaginary part of an analytic function $\varepsilon(\omega)$ from its real part and vice versa.

% In this case, the relations are as follows:
%$$\varepsilon_1(\omega) = {1 \over \pi} \mathcal{P}\!\!\!\int \limits_{-\infty}^\infty {\varepsilon_2(\omega') \over \omega' - \omega}\,d\omega'$$
%and
%$$\varepsilon_2(\omega) = -{1 \over \pi} \mathcal{P}\!\!\!\int \limits_{-\infty}^\infty {\varepsilon_1(\omega') \over \omega' - \omega}\,d\omega',$$
% where $\varepsilon(\omega) = \varepsilon_1(\omega) + i \varepsilon_2(\omega)$  and $\mathcal{P}$ denotes the Cauchy principal value. 

To recover the phase from the amplitude, the function should  be written as: $log(\varepsilon(\omega))=log(\rho(\omega))+i\Theta(\omega)$ with $\rho(\omega)$ its amplitude and $\Theta(\omega)$ its phase. 
The Kramers-Kronig relations can then be applied as follows:
$$\Theta(\omega_0)  =  \frac{2\omega_0}{\pi} \textit{P}\int^{+ \infty}_{0}\frac{ln(\rho(\omega) )}{\omega_0^2-\omega^2}d\omega$$
The basis of this relationship are the Cauchy-Riemann conditions (analyticity of function).  %In this case, the value spectrum can gain value at [0, $\infty$).\par

In some cases this phase can also be obtained simply by using the Hilbert transform of the spectrum:
$$\Theta(\omega_0)  =  -\frac{1}{\pi} \textit{P}\int^{+ \infty}_{- \infty}\frac{ln(\rho(\omega))}{\omega_0-\omega}d\omega.$$

We have implemented in Matlab these two different phase reconstruction methods. The Hilbert transform method has the advantage of being directly available in Matlab, allowing a much faster computing.

\section{description of the simulations}

To test the performance of these methods we have created a small Monte-Carlo program that randomly simulates profiles (${\cal G} (x)$) made of the combination of 5 gaussians according to the formula $ {\cal G} (x)= \sum_{i=1}^{5}  A_i  \exp{\frac{-(\frac{x}{mX} - \mu_i)^2 }{2 \sigma^2_i}} $ where $mX=65536$ and $A_i$, $\mu_i$ and $\sigma_i$ are random numbers with  $x \in [1;mX]$, $A_i \in [0;1] $, $\mu_i \in 0.5 + [ -7.5  ; +7.5  ] \times 10^{-4}/mX $ and  $\sigma_i \in [3;9] \times 10^{-9}$ . The values of these ranges have been chosen to give different profiles without creating disconnected profiles. We have checked how our conclusions are changed outside this range.

Using this formula we have generated 1000 profiles, we then took the absolute value of their Fourier transform $ {\cal F} = \| \mbox{FFT} \left( {\cal G}\right) \|$ and  sampled at a limited number of frequency points ($F_i = {\cal F}(\omega_i)$) as would be done with a real experiment in which the number of measurement points is limited (limited number of detectors or limited number of scanning steps).

Different distributions have been used for the frequencies $\omega_i$: linear, logarithmic, similar to the E-203 experiment at FACET~\cite{E203prstab},... In most sampling schemes 33 frequencies were used to make it comparable with E-203.
%This would have to be detailed in a longer paper.

\begin{figure}[!htb]
 \centering
  \includegraphics*[width=65mm]{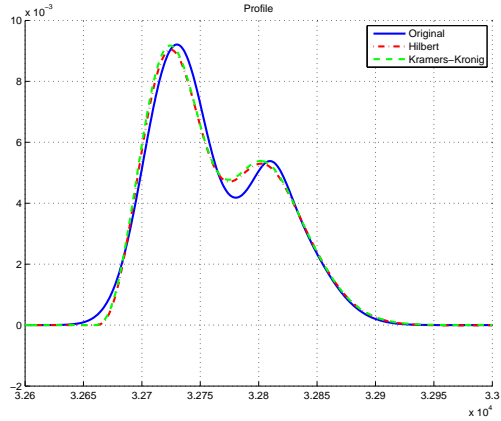}
  \caption{Example of well reconstructed profile. The original profile is in blue and the profiles reconstructed with the Hilbert transform and the full Kramers-Kroning procedures are in red and green respectively.}
   \label{good_profiles}
\end{figure}

\begin{figure}[!htb]
 \centering
  \includegraphics*[width=65mm]{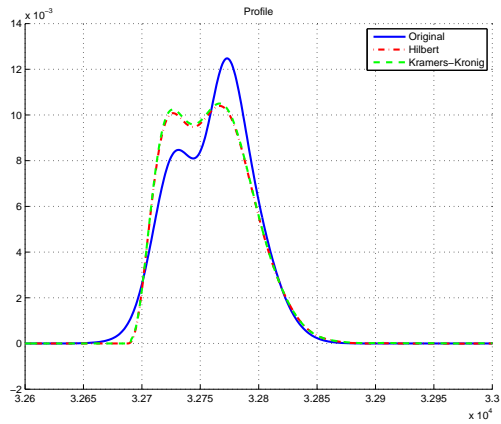}
  \caption{Example of poorly reconstructed profile.  The original profile is in blue and the profiles reconstructed with the Hilbert transform and the full Kramers-Kroning procedures are in red and green respectively.}
   \label{bad_profiles}
\end{figure}

Then, using only these sampled values we applied our reconstruction techniques to reconstruct the original profile.
As can be expected in some cases the reconstruction went very well and in some other cases it was not as convincing. An example of a well reconstructed profile is shown on figure~\ref{good_profiles} and examples of poorly reconstructed profile is shown on figure~\ref{bad_profiles}.

\section{Study of the reconstruction performance}

To estimate the performance of the reconstruction several estimators are available. The $\chi^2$ is one possibility however for two very similar profile but slightly offset, a bad $\chi^2$ will be returned. So we decided to also look at the FWHM which we generalized as FWXM where $X \in [0.1 ; 0.9]$ is the fraction of the maximum value at which we calculate the full width of the reconstructed profile. Here two profiles that are similar but slightly offset (in position or amplitude) will nevertheless return good values (and this is what we want). We have created an estimator $\Delta_{FWXM}$ defined as follow:

$$
\Delta_{FWXM} = \mbox{Max}_{X \in \mbox{rset} }\left| \frac{FWXM_{\mbox{orig}} - FWXM_{\mbox{reco}}}{FWXM_{\mbox{orig}} }\right| 
$$
where $\mbox{rset} = \{ 0.1 ; 0.2 ; 0.5 ; 0.8 ; 0.9\}$, $FWXM_{\mbox{orig}}$ and $FWXM_{\mbox{reco}}$ are the FWXM of the original and reconstructed profiles respectively.

\begin{figure}[!htb]
 \centering
  \includegraphics*[width=70mm]{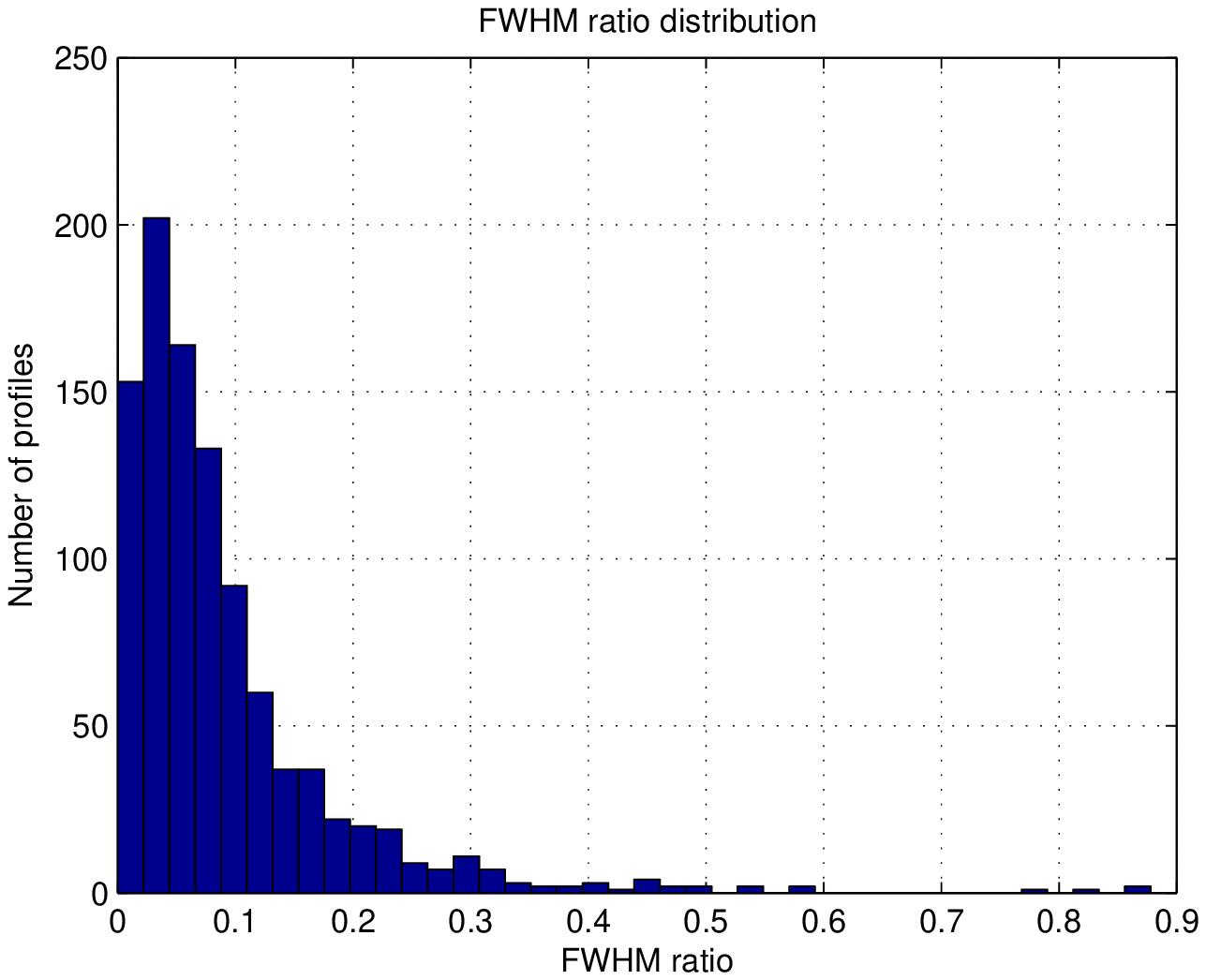} \\
  \includegraphics*[width=70mm]{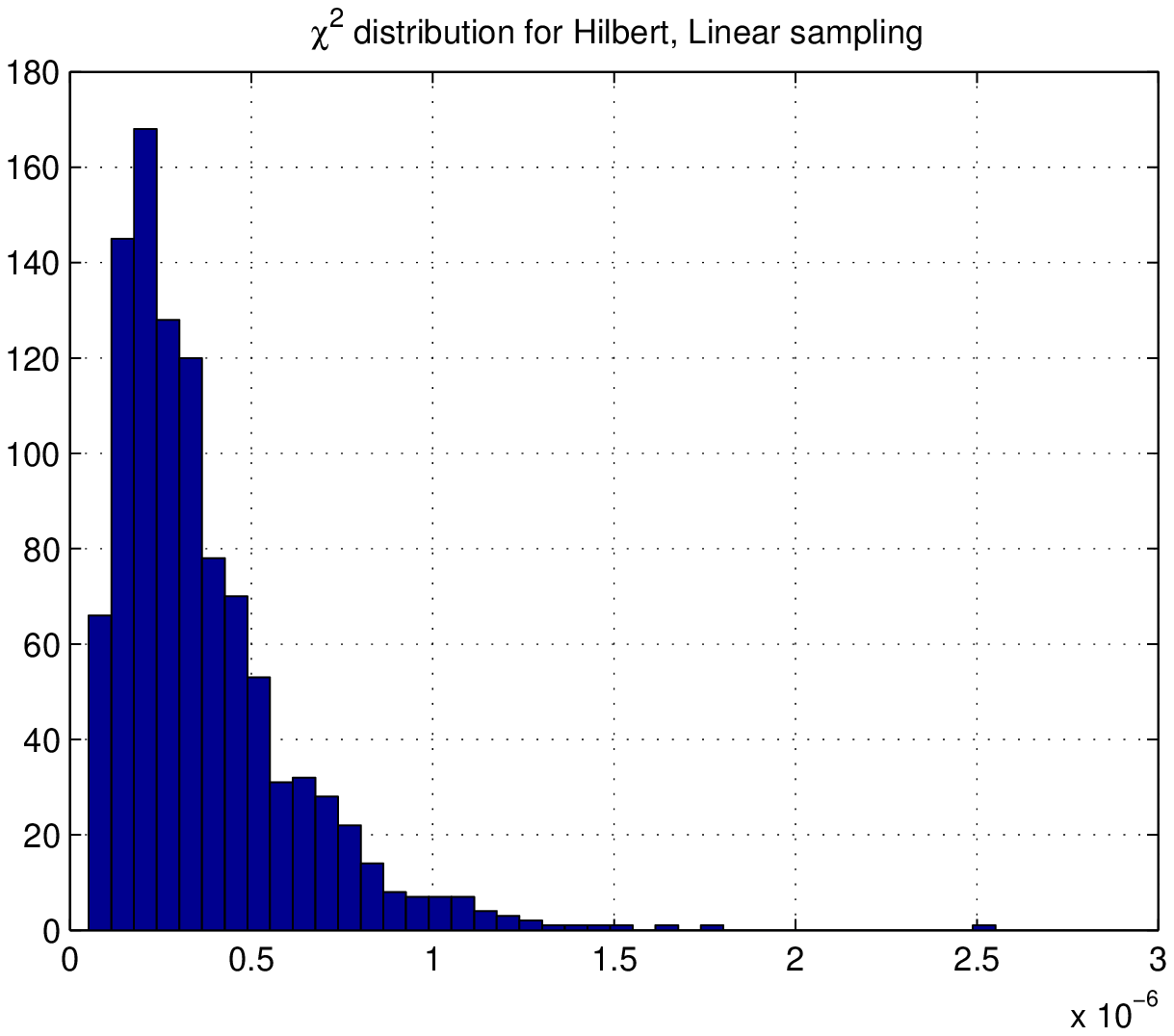} 
  \caption{$\Delta_{FWXM}$  (top) and $\chi^2$ (bottom) distribution of our 1000 simulations reconstructed using the Hilbert transform method.}
   \label{profiles_stats_hilbert}
\end{figure}

\begin{figure}[!htb]
 \centering
  \includegraphics*[width=70mm]{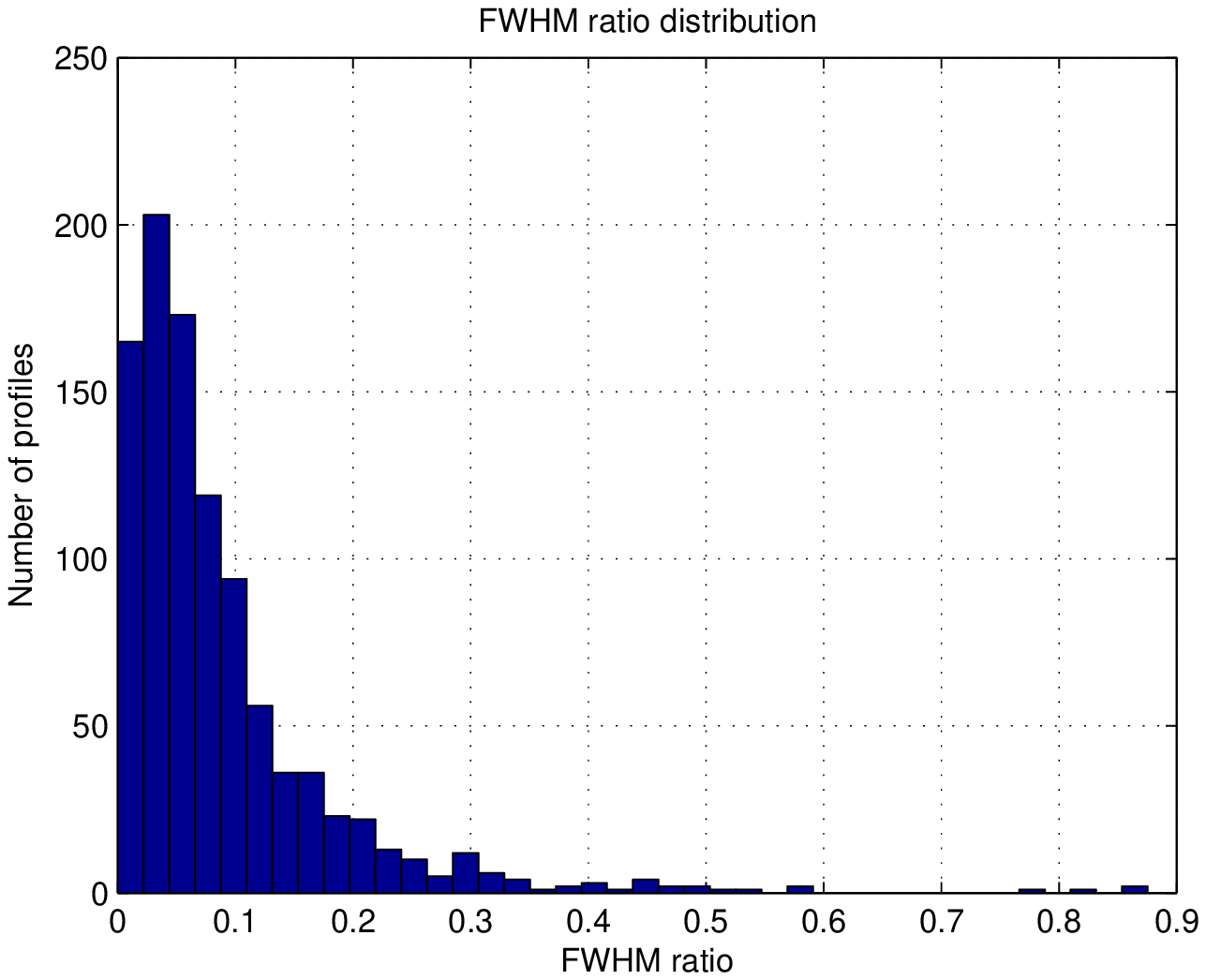} \\
  \includegraphics*[width=70mm]{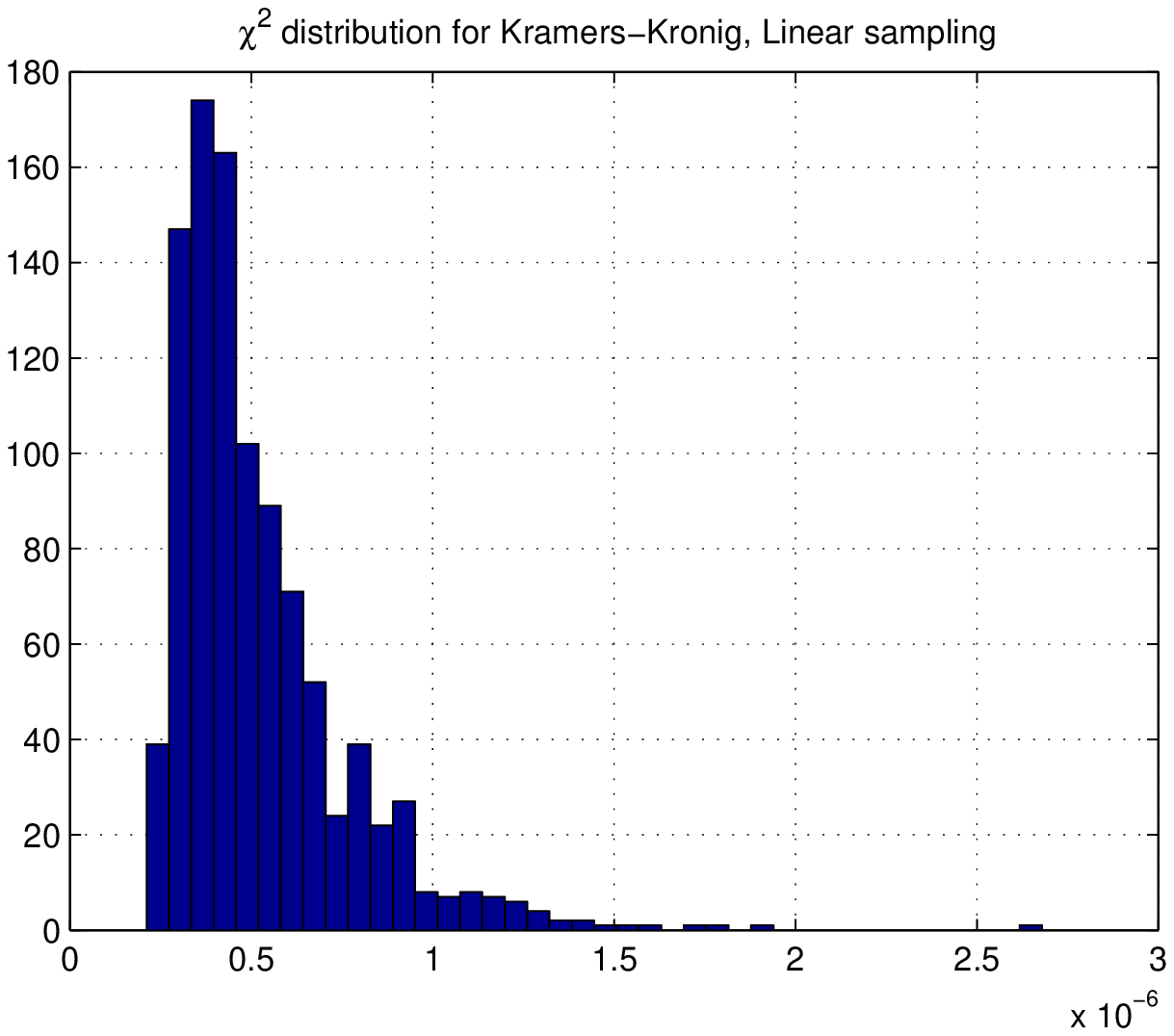} 
  \caption{$\Delta_{FWXM}$   (top)  and $\chi^2$   (bottom)  distribution of our 1000 simulations reconstructed using the Kramers-Kronig reconstruction method.}
   \label{profiles_stats_KK}
\end{figure}

The  $\Delta_{FWXM}$ and  $\chi^2$ distribution of the 1000 simulations which we made and then reconstructed using the Hilbert transform method are shown in figure~\ref{profiles_stats_hilbert} and figure~\ref{profiles_stats_KK} shows the same for a Kramers-Kornig reconstruction. As we can see there is a good agreement.

\begin{figure}[!htb]
 \centering
  \includegraphics*[width=70mm]{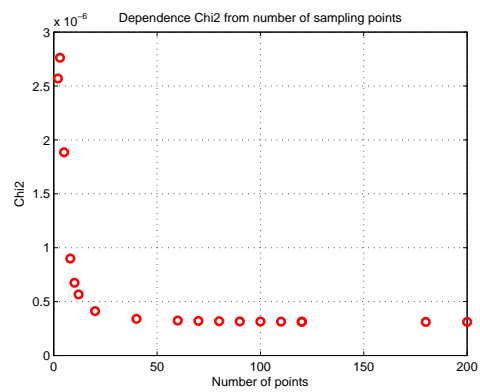}
  \caption{Effect of the sampling frequencies on the $\chi^2$.}
   \label{sampling_chi2}
\end{figure}

The choice of 33 frequencies for the sampling of the spectrum was made to match the current layout used on E-203. However it is important to check if there is an optimum value. Using the same simulations we used different sampling ranging from 5 to 120 sampling frequencies. The effect of changing the sampling frequencies on the  $\chi^2$ is shown on figure~\ref{sampling_chi2}.

% Effect of noise.
% Another important point to check is the effect of the noise on the 

\begin{figure}[!htb]
 \centering
  \includegraphics*[width=70mm]{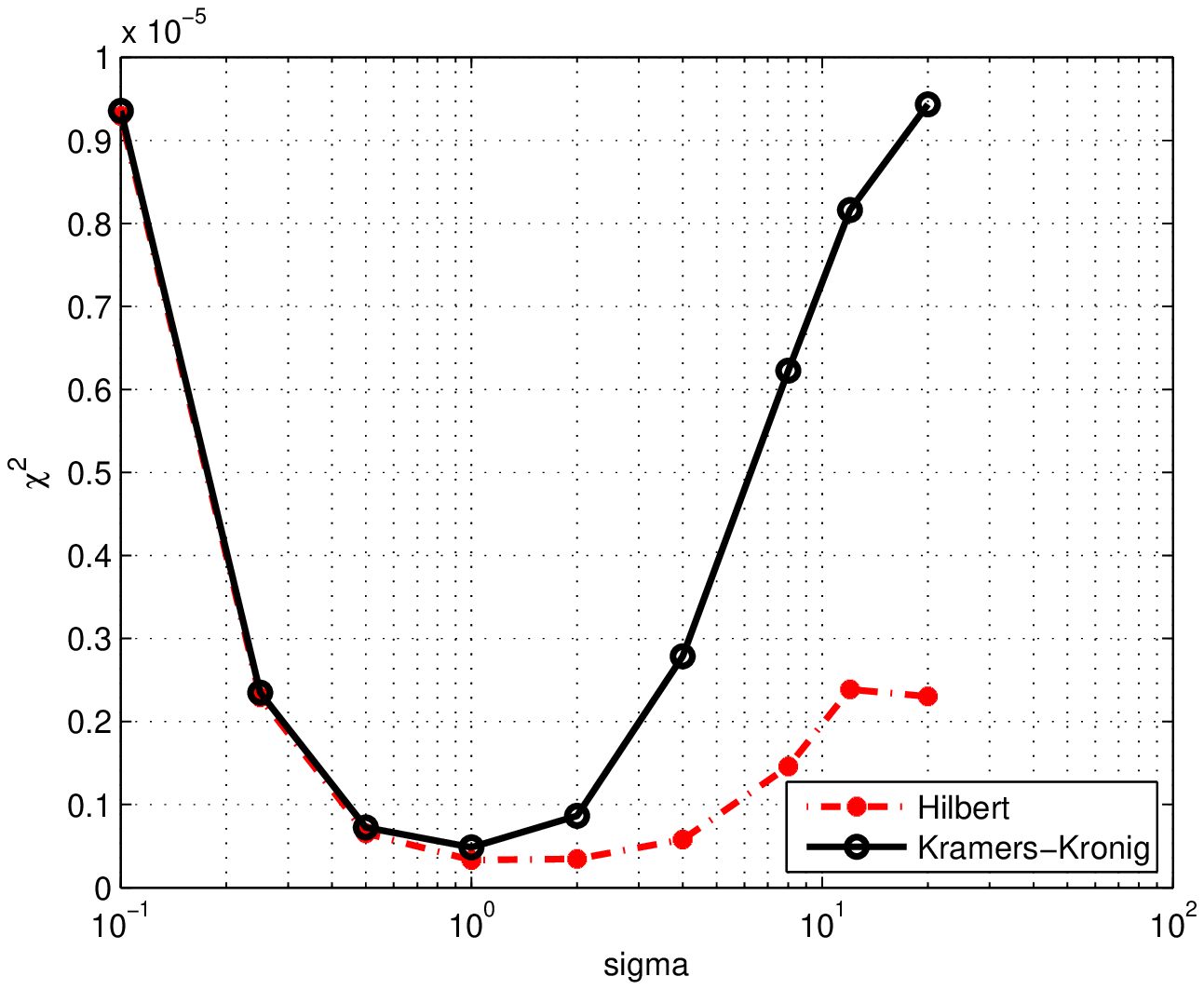}
  \caption{Effect of scaling the constraints on the parameters $\sigma_i$  on the $\chi^2$.}
   \label{sigma_chi2}
\end{figure}

We have also checked how the choice of the constraints on $\sigma_i$ affects the accuracy of the reconstruction. This effect on $\chi^2$ is shown on figure~\ref{sigma_chi2}.

\begin{figure}[!htb]
 \centering
  \includegraphics*[width=70mm]{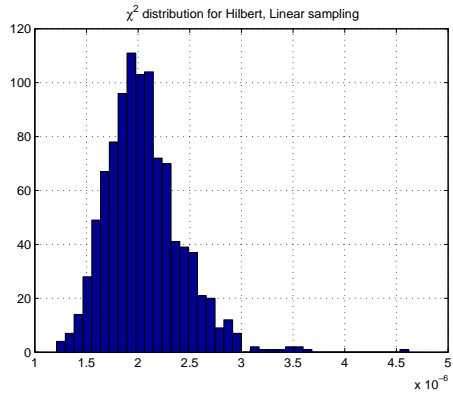} \\
  \caption{Distribution of the $\chi^2$ in the case of a lorenzian distribution.}
   \label{lorenz}
\end{figure}

While doing this work we also became aware of the discussion in~\cite{Pelliccia:2014vba} where it is argued that these reconstruction method have more difficulties with lorentzian profiles rather than gaussian profiles. Therefore we also simulated 1000 lorenzian profiles and performed a similar study. This is shown on figure~\ref{lorenz}. Although the $\chi^2$ is slightly worse in that case than in the case of gaussian profiles we still have a good agreement between the original and reconstructed profiles.

\section{Discussion}

We have performed extensive simulation to estimate the performance of two phase recovery methods in the case of multi-gaussian and lorenzian profiles. In both cases we find that when the sampling frequencies are chosen correctly  we obtain a good agreement between the original and reconstructed profiles (in most cases $\Delta_{FWXM} < 10\%$;  $\chi^2 ~ 10^{-6}$). This confirms that such methods are suitable to reconstruct the longitudinal profiles measured at particle accelerators using radiative methods.

%\bibliographystyle{unsrt}
%\bibliography{biblio}

%%\begin{thebibliography}{9}   % Use for  1-9  references
%\begin{thebibliography}{99} % Use for 10-99 references

%\bibitem{accelconf-ref}
%	C. Petit-Jean-Genaz and J. Poole,
%	``JACoW, A service to the Accelerator Community,''
%	EPAC'04, Lucerne, July 2004, THZCH03,  p.~249,
%	\url{http://www.JACoW.org/e04/papers/THZCH03.PDF}

% \end{thebibliography}

\end{document}